\documentclass{epl}
\usepackage{latexsym}

\title{Joint reality and temporal Bell inequalities}

\shorttitle{Realism and temporal Bell inequalities}

\author{R. Lapiedra\thanks{E-mail address: ramon.lapiedra@uv.es (R. Lapiedra). Telephone: 96 3543077}}

\institute{Department of Astronomy and Astrophysics, University of Valencia, 46100
Burjassot (Valencia), Spain}

\pacs{03.65.Ud}{Entanglement and quantum nonlocality.}
\pacs{03.65.Ta}{Foundations of Quantum Mechanics, measurement theory.}

\begin{document}

\maketitle

\begin{abstract}
Some new temporal Bell inequalities are deduced under joint realism
assumption, using some perfect correlation property. No locality condition
is needed. When the measured system is a macroscopic system, joint realism
assumption substitutes the non-invasive measurability hypothesis
advantageously\textbf{, }provided that the system satisfies the perfect
correlation property. The new inequalities are violated quantically. This
violation can be more severe than the similar violation in the case of
precedent temporal Bell inequalities. Some microscopic and mesoscopic
situations in which these inequalities could be tested are roughly
considered.
\end{abstract}

\section{1. Introduction}

Besides the ordinary Bell inequalities [1, 2] for entangled
systems, there also exist so-called temporal Bell inequalities
[3-5] for a single system. In the seminal paper of Leggett and
Garg [6], the authors consider a macroscopic system and make two
general assumptions:

i) Macroscopic realism: ``A macroscopic system with two or more
macroscopically distinct states available to it will at all times be in one
or the other of these states''.

ii) Noninvasive measurability (NIM): ``It is possible, in principle, to
determine the state of the system with arbitrarily small perturbation in its
subsequent dynamics''.

With these two assumptions, these authors prove some temporal Bell
inequalities for such a macroscopic system, where the measurement times,
\textit{t}$_{i}$, play the role of the polarizer settings in the ordinary
Bell inequalities\textbf{. }NIM assumption is obviously not valid
for quantum systems, and has been criticized for macroscopic systems [7]. In
spite of these criticisms, it seems that the idea of an \textit{ideal
negative} \textit{experiment}, or alternatively the \textit{coupling of the
system to a microscopic probe}, as explained in [6], can change NIM into a
reasonable assumption.

Whatever it be, the main purpose of the present paper is to prove some new
temporal Bell inequalities retaining the realism of assumption i), but
changing the NIM for a new assumption, that encompass the above assumption
i), and becomes extremely natural and plausible if NIM is assumed, but not
necessarily the reverse way. We will call this new assumption the
\textit{joint reality} assumption and we will state it below.

The new temporal Bell inequalities, which apply to any macroscopic
or microscopic system, will be valid provided that the above
assumption holds and the system obeys a so called perfect
correlation property. This property is always valid for any
quantum system and could be valid for macroscopic systems. In any
case, for these last systems, one can always test whether the
property is actually satisfied or not.This perfect correlation
property will be properly described below. Then, it will become
clear that its validity is a first indication of the possible
correctness of the original NIM assumption. The new temporal Bell
inequalities that we will deduce, will be violated by quantum
mechanics more severely than the old temporal Bell inequalities
are.

We will now go on to state this \textit{joint reality} assumption, which
will substitute the above two assumptions i) and ii). Consider an
ensemble of systems \textit{S}, prepared in some way at an initial time.
\textit{S }can be either macroscopic or not, and has a dichotomic magnitude,
\textit{M}, that is, a magnitude which only takes two values, say $\pm $1.
We will measure \textit{M} for three different values,\textit{ a},
\textit{b}, and \textit{c}, of an external parameter. (An example could be a
one half spin particle measured on three different directions). Then, the
above joint reality assumption assumes the joint existence of a reality
behind any obtainable measurement outcome. More precisely\textbf{,
w}e will denote this joint reality by \textit{(a}$^{\alpha
}$\textit{b}$^{\beta }$\textit{c}$^{\gamma }$\textit{)}, where
\textit{$\alpha $, $\beta $, $\gamma =\pm $}. That is,
\textit{(a}$^{\alpha }$\textit{b}$^{\beta }$\textit{c}$^{\gamma }$\textit{)
}is the reality such that, if we took a measurement above, for the parameter
value\textit{ a}, we would obtain\textit{ $\alpha $1}, i.e.,
\textit{+1} or --\textit{1}, according to the value of
\textit{$\alpha $}. Similar for the other \textit{directions b} and
\textit{c}. Notice that, like d'Espagnat in Ref. [9], we assume the
existence of a reality for all the possible results of all possible
measurements, even if each actual measurement is taken in a single randomly
chosen direction. As we have commented above, this kind of reality is a very
natural assumption as far as one assumes NIM. We will comment below why it
will be also interesting to assume initially this kind of reality in the
case of a quantum system, where obviously the NIM assumption is non valid.

On the other hand, suppose we perform two immediately consecutive
measurements for the same external parameter value. Then, we expect that the
corresponding outcomes are \textit{perfectly correlated}, i. e., if
the first measurement value is +1, the second one is always +1, and likewise
for a value of --1. This will be called the \textit{perfect correlation
}property. Obviously, this property is always satisfied when \textit{S} is
a quantum system\textbf{,} a \textit{qubit }in this paper\textit{
}(i.e., a quantum system whose space of states is 2-dimensional), and we
take pairs of immediately consecutive measurements for different
\textit{directions} randomly chosen among the same three different external
parameter values\textbf{. }For a macroscopic system, perfect correlation
property can be expected to be valid as far as the NIM assumption is
correct. But, here, the question is simply whether experience will show or
not that the property is satisfied. If it does, we meet the conditions to
prove our temporal inequalities. Otherwise, we could not prove them.

Thus, under joint realism assumption, using the perfect
correlation\textbf{ }property, we will prove some new temporal
Bell inequalities where the measurement times, \textit{t}$_{i}$,
of Ref. [6] will be replaced by the above parameter values,
\textit{a},\textit{ b}, and\textit{ c} (in spite of this
replacement, we will always call temporal Bell inequalities the
new inequalities to be obtained here). In Ref. [8] a similar
problem is addressed. The authors proved the temporal inequalities
which are analogous to the ordinary CHSH inequalities$^{ }$[2],
under the ``locality in time'' assumption. We will not use here
this doubtful assumption, which will be replaced by the above
joint reality assumption and the use of the perfect correlation
property.

\section{2. Proving the new temporal Bell inequalities}

Let us consider the above perfectly correlated system, \textit{S},
with its dichotomic magnitude, \textit{M}, measured randomly for
the external parameter values \textit{a}, \textit{b}, and
\textit{c}. We want to prove a temporal Bell inequality for the
outcomes of these measurements, assuming joint realism and using
perfect correlation. In order to do so, we will adapt to our case
a proof of an ordinary Bell inequality for a singlet-state pair of
entangled qubits. Here we adapt this proof in the form given by
d'Espagnat [9], even if the proof was first given by Wigner [10].

Let us be more precise about the kind of experiment we are going to
consider. In each system, \textit{S}, of the above ensemble, we take two
immediately consecutive measurements of \textit{M} for two independent
values, randomly chosen, of the three fixed external parameter values
\textit{a}, \textit{b}, and \textit{c}. We will call each of these pairs of
measurements a \textit{run}. Then, as we have explained before, we assume
the existence of a joint reality behind any obtainable measurement outcome.
$^{\alpha \gamma }$\textit{$\beta $}$^{\alpha \gamma }$\textit{$\alpha $}et
us consider the number of these joint realities \textit{(a}$^{\alpha
}$\textit{b}$^{\beta }$\textit{c}$^{\gamma }$\textit{)}, which are
present \textit{after the first} \textit{measurement of every run
and before the second measurement}. Let us denote this number by
\textit{N(a}$^{\alpha }$\textit{b}$^{\beta }$\textit{c}$^{\gamma
}$\textit{)}. We will define

\textit{N(a}$^{+}$\textit{b}$^{-}$\textit{) $\equiv $
N(a}$^{+}$\textit{b}$^{-}$\textit{c}$^{+}$\textit{)}$_{ }$\textit{+
N(a}$^{+}$\textit{b}$^{-}$\textit{c}$^{-}$\textit{), } (1) $_{ }$

\textit{N(a}$^{+}$\textit{c}$^{-}$\textit{) $\equiv $
N(a}$^{+}$\textit{b}$^{+}$\textit{c}$^{-}$\textit{)}$_{ }$\textit{+
N(a}$^{+}$\textit{b}$^{-}$\textit{c}$^{-}$\textit{),}$_{ }$ (2)

\textit{N(b}$^{+}$\textit{c}$^{-}$\textit{) $\equiv $
N(a}$^{+}$\textit{b}$^{+}$\textit{c}$^{-}$\textit{)}$_{ }$\textit{+
N(a}$^{-}$\textit{b}$^{+}$\textit{c}$^{-}$\textit{). } (3)

From this, we readily have:

\textit{N(a}$^{+}$\textit{c}$^{-}$\textit{) $\le $
N(a}$^{+}$\textit{b}$^{-}$\textit{) + N(b}$^{+}$\textit{c}$^{-}$\textit{). }
(4)

Now, let us consider, for example, the number of \textit{runs},
\textit{N[a}$^{+}$\textit{,b}$^{-}$\textit{]}, where
\textit{a}$^{+}$ is the outcome of the first measurement and
\textit{b}$^{-}$ the outcome of the second one. (Notice that we use square
brackets for measurement outcomes and standard brackets for hypothetical
realities). Obviously, these \textit{runs} can only come from the above
realities \textit{(b}$^{-}$\textit{)} between the first and the
second measurement. Furthermore, from the perfect correlation assumption,
they can only come from the more specific realities
\textit{(a}$^{+}$\textit{b}$^{-}$\textit{)}. (The notation
\textit{(b}$^{-}$\textit{)} and \textit{(a}$^{+}$\textit{b}$^{-}$\textit{)}
should be obvious). Then, given a reality between both
measurements\textbf{,} such as
\textit{(a}$^{+}$\textit{b}$^{-}$\textit{)},$_{ }$what is the probability of
obtaining a\textit{ run} like \textit{[a}$^{+}$\textit{,b}$^{-}$\textit{]}?
Since the choice of any one of the three parameters, \textit{a}, \textit{b},
\textit{c}, is a random choice, this probability is just \textit{1/9}. This
means that we can write

\textit{N(a}$^{+}$\textit{b}$^{-}$\textit{) =
9N[a}$^{+}$\textit{,b}$^{-}$\textit{], } (5)

and similarly for \textit{N(a}$^{+}$\textit{c}$^{-}$\textit{)} and\textit{
N(b}$^{+}$\textit{c}$^{-}$\textit{).} Thus, taking into account Eq. (4), we
obtain the temporal Bell inequality:

\textit{N[a}$^{+}$\textit{,c}$^{-}$\textit{] $\le $
N[a}$^{+}$\textit{,b}$^{-}$\textit{] +
N[b}$^{+}$\textit{,c}$^{-}$\textit{],} (6)

for the observable quantities
\textit{N[a}$^{+}$\textit{,c}$^{-}$\textit{]},\textit{
N[a}$^{+}$\textit{,b}$^{-}$\textit{]}, and\textit{
N[b}$^{+}$\textit{,c}$^{-}$\textit{]. }

Notice that in this proof it is essential to define the above
joint reality \textit{(a}$^{\alpha }$\textit{b}$^{\beta
}$\textit{c}$^{\gamma }$\textit{) }as the joint reality which is
present before the second measurement of the run and after the
first one. In this way, the reality can only be changed by the
second measurement. But this change is irrelevant to the
completion of our proof, since in a \textit{run} we do not
consider a third measurement.

If one prefers to speak in terms of probabilities corresponding to the
numbers in inequality (6), we can write this inequality as

\textit{P(a}$^{+}$\textit{,c}$^{-}$\textit{) $\le $
P(a}$^{+}$\textit{,b}$^{-}$\textit{)+P(b}$^{+}$\textit{,c}$^{-}$\textit{),}
(7)

and in a similar way

\textit{P(a}$^{-}$\textit{,c}$^{+}$\textit{) $\le $
P(a}$^{-}$\textit{,b}$^{+}$\textit{)+P(b}$^{-}$\textit{,c}$^{+}$\textit{).}
(8)

Or, in terms of the expected value,

\textit{E(a,b) =
P(a}$^{+}$\textit{,b}$^{+}$\textit{)+P(a}$^{-}$\textit{,b}$^{-}$\textit{)--P(a}$^{+}$\textit{,b}$^{-}$\textit{)--P(a}$^{-}$\textit{,b}$^{+}$\textit{)},
(9)

taking into account inequalities (7) and (8), we obtain:

\textit{E(a,b)+E(b,c)--E(a,c) $\le $ 1}.  (10)

At first sight, one might think that inequalities (7), (8), or
(10) are of no interest since, if they were experimentally
violated, this could always be explained by some transmission of
information between the two consecutive measurements of the
\textit{run}. But this is not true since, as we have seen,
inequalities (7), (8), and (10) have been deduced from the joint
realism assumption\textbf{, }using the perfect correlation
property, without any further assumptions. Therefore, we can
transmit all kinds of information we want between both
measurements, but if perfect correlation and joint realism are
preserved, as we assume, inequalities (7), (8), and (10) must
remain true.

Now, we could find that the joint realism assumed in the present paper is a
too restrictive postulate in the case of a quantum system, and, thus, a non
convincing postulate for such a system. In fact, in quantum mechanics, two
non commuting observables cannot be measured at the same time. Furthermore,
the orthodox interpretation of the theory assumes that it is not only that
we cannot jointly measure them, but it asserts that the corresponding joint
reality does not exist. On the other hand, as we will see in the next
Section, quantum mechanics entails the violation of the present temporal
Bell inequalities. Then, we can say now that the non existence of joint
reality in the case of a qubit is no more a question of interpretation, but
a prediction of quantum mechanics which could be easily tested
experimentally, by testing these temporal Bell inequalities. Obviously, it
is to be expected that experience will agree in this point with quantum
mechanics and so that it will reject the assumed joint reality.

\section{3. Quantum violation of the temporal Bell inequalities }

Let us assume that our system \textit{S} is a qubit. Then, a normalized
general state, /\textit{$\psi >$}, can be written as:

\textit{/$\psi >$ = s /e+$>$ + (1--
s}$^{2}$\textit{)}$^{1/2}$\textit{ e}$^{i\phi }$\textit{ /e--$>$},
(11)

where \textit{ /e+$>$ }and \textit{ /e--$>$ }are the eigenstates of
eigenvalues $\pm $1, respectively, for a given ``direction'' \textit{e}.
Since for any ``direction'' \textit{x }the corresponding eigenstates,
\textit{/x+$>$ }and \textit{ /x--$>$}, are orthogonal unit vectors in a
2-dimensional Hilbert space, it is straightforward to show that an angle
\textit{$\alpha $}$_{x}$ always exists such that

\textit{/x+$>$ =[(1+cos $\alpha
$}$_{x}$\textit{)/2]}$^{1/2}$\textit{/e+$>$ + [(1--cos $\alpha
$}$_{x}$\textit{)/2]}$^{1/2}$\textit{/e--$>$, } (12)

\textit{/x--$>$ =[(1--cos $\alpha
$}$_{x}$\textit{)/2]}$^{1/2}$\textit{/e+$>$ -- [(1+cos $\alpha
$}$_{x}$\textit{)/2]}$^{1/2}$\textit{/e--$>$. } (13)

This means, as it is well-known, that \textit{x} and \textit{e} can always
be interpreted as two unit 3-vectors in \textbf{\textit{R}}$^{3}$,
\textbf{\textit{x }}and \textbf{\textit{e}}, respectively, which appear in
these equations only through their 3-scalar product
\textbf{\textit{x.e}}\textit{ = cos $\alpha $}$_{x}$.

Hence, when measuring the above dichotomic magnitude \textit{M} for the
three external parameter values, \textit{a}, \textit{b}, and \textit{c}, we
can always say that these measurements have been taken for the corresponding
unit 3-vectors, \textbf{\textit{a}},\textbf{\textit{ b}},
and\textbf{\textit{ c}}.

Let us consider the different probabilities,
\textit{P(}\textbf{\textit{a}}$^{\pm }$\textit{,}\textbf{\textit{b}}$^{\pm
}$\textit{)}\textbf{, }of obtaining $\pm $1 for the two consecutive
measurements of the \textit{runs} where chance has selected, respectively,
the unit 3-vectors \textbf{\textit{a }}and\textbf{\textit{ b}}. After some
basic algebra, we find

\textit{P(}\textbf{\textit{a}}$^{+}$\textit{,}\textbf{\textit{b}}$^{-}$\textit{)
= s}$^{2}$\textit{(1--}\textbf{\textit{a.b}}\textit{) /2}\textbf{,
}\textit{P(}\textbf{\textit{a}}$^{-}$\textit{,}\textbf{\textit{b}}$^{+}$\textit{)
= (1--s}$^{2}$\textit{)(1--}\textbf{\textit{a.b}}\textit{) /2}.
(14)

Thus, according to Eq. (9), we obtain:

\textit{E(}\textbf{\textit{a}}\textit{,}\textbf{\textit{b}}\textit{)
= }\textbf{\textit{a.b}}\textit{, } (15)

which differs in sign from the similar result for the expected
value in the case of an entangled pair of qubits in the singlet
state. (Obviously, for \textit{E(a,c) }and\textit{ E(c,b)}, we
have similar equations to Eq. (15)).

Notice that the result (15) has the remarkable property of being
independent of the initial state of the particle [8], that is, in
(15),
\textit{E(}\textbf{\textit{a}}\textit{,}\textbf{\textit{b}}\textit{)
}does not depend on \textit{s} or \textit{$\phi $ }appearing in
Eq. (12), while \textit{P(}\textbf{\textit{a}}$^{\pm
}$\textit{,}\textbf{\textit{b}}$^{\pm }$\textit{)} does depend on
\textit{s}. All this \textbf{means that the version (10)}
\textbf{of our temporal Bell inequalities} does not depend on the
initial state of the system \textit{S}\textbf{, }while versions
(7) or (8) do\textbf{.}

Bearing in mind Eq. (15) and the similar ones, the Bell inequality
(10) becomes

\textbf{\textit{a.}}(\textbf{\textit{b}}\textit{--}\textbf{\textit{c}}\textit{)+}\textbf{\textit{b.c}}\textit{
$\le $ 1, } (16)

which is maximally violated by any two orthogonal unit 3-vectors
\textbf{\textit{b}} and \textbf{\textit{c}}, if the unit 3-vector
\textbf{\textit{a}} is collinear to
\textbf{\textit{b}}\textit{--}\textbf{\textit{c}}. In all these
cases the left hand side of inequality (16) takes the value $\surd
$2.

Similarly one can see that the quantum mechanics of qubit (11)
violates inequalities (7) or (8), but this violation depends on
the initial state of the qubit. For example, if this initial state
is the eigenstate \textit{$/$a+$>$}, for the different
probabilities appearing in inequality (7) one finds:

\textit{P(a}$^{+}$\textit{,c}$^{-}$\textit{) =
(1--}\textbf{\textit{a.c}}\textit{)/2,
P(a}$^{+}$\textit{,b}$^{-}$\textit{) =
(1--}\textbf{\textit{a.b}}\textit{)/2,
P(b}$^{+}$\textit{,c}$^{-}$\textit{)
=(1+}\textbf{\textit{a.b}}\textit{)(1--}\textbf{\textit{b.c}}\textit{)/4.
} (17)

Then, for inequality (7) we get:

\textbf{\textit{b.}}(\textbf{\textit{a}}\textit{+}\textbf{\textit{c}}\textit{)--2}\textbf{\textit{a.c+}}\textit{(}\textbf{\textit{a.b}}\textit{)(}\textbf{\textit{
b.c}}\textit{) $\le $ 1.} (18)

For \textbf{\textit{a.c}}\textit{ = 0} and
\textbf{\textit{b}}\textit{ =
(}\textbf{\textit{a}}\textit{+}\textbf{\textit{c}}\textit{)/$\surd
$2}, this inequality is more severely violated than the above
inequality (16), since the left hand side becomes \textit{$\surd
$2+1/2 }for this configuration, instead of the above
\textit{$\surd $2} for inequality (16) .

\section{4. Some examples}

Once we have seen that the temporal Bell inequalities (7), (8),
and (10) can be violated by quantum mechanics, we roughly turn to
the question of how this violation could be experimentally
produced. Here, the problem is that we need to perform two
successive measurements on the same system, and not merely in two
different parts of the same system, as in the ordinary space
entangled Bell inequalities. Then, we must guarantee that the
first of these two measurements is always a first class
measurement, that is, a preparation-like measurement, in order to
preserve the existence of the system and then be able to take the
second measurement. These conditions can be fulfilled, in
principle, in the case where the measured system is a one half
spin particle, whose spin is successively measured in different
directions, with a Stern-Gerlach device. We must then distinguish
two cases, according to whether we want to test inequality (10),
or alternatively one of the two inequalities (7) or (8).

In the first case, in order to obtain the measurement outcomes to
calculate, for example, the expected value \textit{E(a,b)} in
(10), we must perform two different measurement series, two run
series to be more precise (following a similar strategy to the one
stated in [6], where the authors combine different ideal
negative-result setups). One run series to obtain the
probabilities \textit{P(a}$^{+}$\textit{,b}$^{+}$\textit{)
}and\textit{ P(a}$^{+}$\textit{,b}$^{-}$\textit{)} and the other
run series to obtain the probabilities
\textit{P(a}$^{-}$\textit{,b}$^{-}$\textit{) }and\textit{
P(a}$^{-}$\textit{,b}$^{+}$\textit{)}. In the first series we only
retain the \textit{a}$^{+}$ outcomes corresponding to a
preparation-like measurement. In this way, the spin particle is
still available for a second measurement in direction
\textit{b}\textbf{.} We will proceed in a similar way for the
second series, where in another large and identical ensemble we
will only retain the \textit{a}$^{-}$ outcomes corresponding to
another preparation-like measurement. In this way, we will be able
to measure the three expected values \textit{E(a,b)},
\textit{E(b,c)} and \textit{E(a,c)}, and thus to test inequality
(1o). Notice that, as we have already remarked, in the present
case we do not need to prepare the system in any particular state
before each run.

If reader feels that the two different series of measurements we
have considered in the above case are unfair, they can consider
the second case, which is the one considered to test, let us say,
inequality (7), and is, furthermore, an interesting case in
itself. In this second case, we assume that the \textit{a}$^{+}$
and \textit{b}$^{+}$ outcomes in (7), corresponding to the first
measurements of each run, refer to preparation-like measurements.
In the present case, the different probabilities which appear in
(7) do depend on the particle state before each run. Then, for
each run, we will prepare this previous state as an
\textit{a}$^{+}$ eigenstate. This is what has been assumed in
order to deduce the inequality (18) that, as we have seen, is more
severely violated than the corresponding inequality (16). In all,
for each run, we must perform three successive Stern-Gerlach
measurements: first, we must prepare the\textit{ a}$^{+}$
eigenstate, and then perform two successive measurements from this
eigenstate. In this way we will be able to measure the three
probabilities of inequality (7) and thus to test this inequality.

In the event that one could overcome the practical difficulties
which might arise in actually carrying all this out, why is it of
interest to do an experiment like the ones we have outlined, let
it be in the case of the half one spin, or in the case of some
other sort of qubit? As we have explained above, the proof of the
temporal Bell inequalities we are considering here, do not rely on
the locality assumption. Instead, the proof only depends on the
joint reality assumption and the observed perfect correlation
property. Thus, if these temporal inequalities were experimentally
violated, only realism would be disregarded, without further
concerns about locality conditions, contrarily to the case of
standard Bell inequalities\textbf{. }Hence, the locality loophole
is not present here. Remember that the existence of this loophole,
in the case of the ordinary Bell inequalities, moved people to
make experiments like those performed by Aspect and other similar
ones [11]. The other well known loophole is the one referred to as
the fair sample loophole (see [12], for example), which is mainly
a consequence of the low detection level of the photons used in
experiments. In spite of this, experimentalists have used visible
photons in their experiments to test ordinary Bell inequalities
because, in practice, it is relatively easy to produce entangled
pairs of such photons. Nevertheless, we do not need an entangled
system of particles to test our temporal Bell inequalities. We
only need, in principle, a single qubit. So, it is to be expected
that any qubit, with a detection level that is high enough, could
be chosen to avoid this second loophole. This would mean, then,
that both loopholes have been closed at the same time, as far as
joint reality is involved.

On the other hand, in [13], the authors consider a micrometer
sized superconducting loop, with Josephson junctions. They have
been able to observe the quantum superposition of two different
macroscopic states of this mesoscopic system. Furthermore, several
authors (see ref. [7]) have considered the possibility of using a
similar experimental device to test the original temporal Bell
inequalities, whose inequalities use three different measurement
times. Perhaps, such an experimental device could also be used to
test any of our temporal Bell inequalities (7), (8), or (10), in
which three external parameter values play the role of those
measurement times. The proposal would be to apply an external
magnetic flux to the loop, lower than the superconducting flux
quanta, randomly chosen among three different fixed fluxes each
time. These three fluxes would now play the role of the above
three ``directions'', \textit{a}, \textit{b} and\textit{ c}. Then,
the dichotomic response of the system would be that one or another
of the two bottom persistent currents of opposite sign would
appear. In the third reference of [7], the authors try to show
that temporal Bell inequalities cannot help to discriminate
realism and quantum mechanics in such a superconducting loop. But
they refer to the original temporal Bell inequalities, in which
the role of the three external parameters, \textit{a}, \textit{b}
and \textit{c}, is played by three different times. It is not
clear to us that these difficulties will be still present in the
case of our temporal Bell inequalities. So, the problem will be
rather to make sure that in our case perfect correlation is
satisfied.

\section{6.-Conclusions}

In the present paper, we have proved some temporal Bell inequalities under
the assumptions of ``joint realism'', using the perfect correlation
property, for any kind of physical system, macroscopic or microscopic, with
a randomly dichotomic magnitude, i.e., a magnitude which randomly takes two
values. The measurement outcomes are the response of the system to some
different external parameter values, as in the standard Bell inequalities.
These different parameter values play the role of the different measurement
times in the seminal paper of Legget and Garg [6]. In this paper, the
authors deal with realism and NIM assumptions in the context of macroscopic
systems. In the present paper, we deal jointly with macroscopic or
microscopic systems, by substituting both assumptions for the joint
reality assumption, and by using the perfect correlation property.
Contrarily to the case of NIM assumption, joint reality can be asserted, in
principle, either for microscopic systems or for macroscopic ones.

On the other hand\textbf{,} the perfect correlation property is always
verified by quantum systems. When the physical system is a macroscopic one,
one must verify whether the perfect correlation property is satisfied. One
can expect that this property will be verified in the macroscopic case on
the grounds of the joint reality assumption\textbf{.}

The new assumption of joint reality substitutes NIM assumption
advantageously, since that assumption is less restrictive than this one, it
can be applied, in principle, to quantum systems, and finally it will
provide us with new temporal Bell inequalities which are more severely
violated by quantum mechanics than the old ones.

The expected values and the probabilities, which appear in the new
temporal Bell inequalities (7) or (8), and (10), have to do with
pairs of immediately consecutive measurements made in the system
as such, and not in different parts of an entangled system, as in
the ordinary Bell inequalities. This is of course well known, but
what is essential in the present paper is that no locality
assumption is needed to prove the present temporal Bell
inequalities. This is essential since here we deal with the two
immediately consecutive measurements of the same \textit{run},
i.e., two consecutive measurements in the same system, and so,
some sort of information transmission between these two
measurements is now present. In spite of this, the locality
loophole is avoided. It is to be expected that the efficiency
loophole can also be avoided in some special cases, and likewise
closing the loophole problem, as far as this new temporal Bell
inequalities are concerned.

In particular, when trying to prove our temporal Bell inequalities for a
given macroscopic system, we need only be sure that the perfect correlation
property holds or nearly holds. We do not need to be concerned with any kind
of information which could be propagated between two successive
measurements.

\acknowledgements
This work has been supported by the Spanish MCyT (Project AYA 2003-08739-C02
partially founded with FEDER) and also by the Generalitat Valenciana (grupo
03/170).

I am grateful to E.Santos for fruitful discussions.

\end{document}